\documentclass[secnumarabic,amssymb,amsmath,aps,prd,nofootinbib]{revtex4}
\usepackage{graphicx}

\newcommand{\rp}{\right}
\newcommand{\lp}{\left}
\newcommand{\beq}{\begin{equation}}
\newcommand{\eeq}{\end{equation}}
\newcommand{\bal}{\begin{align}}
\newcommand{\eal}{\end{align}}

\newcommand{\fl}{{({\rm flat})}}
\newcommand{\s}{k}
\newcommand{\q}{n}

\newcommand{\eff}{{\rm eff}}
\newcommand{\mmu}{\nu}
\newcommand{\mbr}{{m}}
\newcommand{\uu}{{\cal U}}
\newcommand{\yy}{y}
\newcommand{\la}{\langle}
\newcommand{\ra}{\rangle}
\newcommand{\tmn}{\la T_{\mu\nu}\ra}
\newcommand{\phisq}{\la \Psi^2 \ra}
\newcommand{\kk}{\textsc{kk}}
\newcommand{\bs}{{bs}}

\begin{document}

\title{Vacuum destabilization from Kaluza-Klein modes \\in an inflating brane}

\author{ Oriol  Pujol{\`a}s\footnote{pujolas@yukawa.kyoto-u.ac.jp} and
Misao Sasaki\footnote{misao@yukawa.kyoto-u.ac.jp}}

\address{Yukawa Institute for Theoretical Physics, Kyoto University, Kyoto
  606-8502, Japan}

\begin{abstract}

We discuss the effects from the Kaluza-Klein modes in the brane
world scenario when an interaction between bulk and brane fields
is included. We focus on the bulk inflaton model, where a bulk
field $\Psi$ drives inflation in an almost $AdS_5$ bulk bounded by
an inflating brane. We couple $\Psi$ to a brane scalar field
$\varphi$ representing matter on the brane. The bulk field $\Psi$
is assumed to have a light mode, whose mass depends on the
expectation value of $\varphi$.
The KK modes form a continuum with masses $m>3H/2$, where $H$ is
the Hubble constant. To estimate their effects, we integrate them
out and obtain the 1-loop effective potential $V_\eff(\varphi)$.
With no tuning of the parameters of the model, the vacuum becomes
(meta)stable -- $V_\eff(\varphi)$ develops a true vacuum at
$\varphi\neq0$. In the true vacuum, the light mode of $\Psi$
becomes heavy, degenerates with the KK modes and decays.
We comment on some implications for the bulk inflaton model.
Also, we clarify some aspects of the
renormalization procedure in the thin wall approximation, and show
that the fluctuations in the bulk and on the brane are closely
related.\\


\hfill \vbox{ \hbox{YITP-05-39} }
\end{abstract}

\maketitle

\section{introduction}

The Brane World (BW) scenario \cite{add,rsI,rsII} has recently
concentrated a lot of attention in cosmology both for its rich
phenomenology and for the link that it makes between string theory
and observation. At present, one of the most challenging issues
for the application of the BW paradigm to cosmology is the
computation of the quantum fluctuations of bulk fields, especially
during inflation since this might leave some signature of the
extra dimensions in observables like the CMB. The new degrees of
freedom brought by the extra dimensions are described in the four
dimensional language by a collection of massive fields, the
Kaluza-Klein (KK) modes. In models with an infinite extra
dimension, the spectrum of KK modes is continuous. If the brane
inflates, the lowest-lying KK mode has a mass of the order of the
Hubble constant $H$ \cite{gasa}, and the role of the KK modes can
be quite important.

A model of this type is the 'bulk inflaton model' \cite{kks,hs},
in which inflation is driven by a bulk scalar field $\Psi$. The
field $\Psi$ is assumed to have a 'bound state' with mass
$m_\bs\ll H$, so that its vacuum fluctuations are large. The
contribution from the KK modes of $\Psi$ to the fluctuations when
the bound state is massless was shown to be negligible even in the
$H\ell\gg1$ limit \cite{kks} (here $\ell$ is the bulk AdS radius),
when higher dimensional effects are important in principle. In
\cite{shs}, it was found that the the KK modes do not contribute
to the spectrum in the $m_\bs\ell\ll1$ limit, whereas in the
opposite limit, the KK contribution is larger but saturates to
less than 10\% for typical values of the parameters \cite{shs}.
This was interpreted as an indication that the bulk inflaton model
is a viable alternative to the four dimensional cosmology (see
also \cite{hts}).

The computation of the KK contribution to the power spectrum of
primordial fluctuations is slightly obscured by divergences
associated to the presence of the infinite tower of massive modes
\cite{shs,kks}. Even though the sensitivity to the cutoff is only
logarithmic, one would desire a physical interpretation of the
cutoff, or a well defined renormalization scheme that justifies
the subtraction of divergences.

The purpose of this article is to compute the quantum fluctuations
$\phisq$ (summed over all wavelengths), for which the
renormalization procedure follows standard methods. This provides
a useful way to handle the higher dimensional effects, especially
when interactions are taken into account. A first approximation to
include the 'heavy' modes is to integrate them out. This refers to
performing explicitly the path integral over these modes, and is
equivalent to computing the 1-loop effective potential. As we
shall see, the quantum effects of the KK modes can play an
important role {\em e.g.} in establishing the vacuum stability of
the model.

We shall consider the bulk inflaton model \cite{kks,hs} in a
slow-roll approximation where the background space is frozen. The
fluctuations of the inflaton are described by a test bulk scalar
$\Psi$ in spacetime of the RSII model \cite{rsII} with one de
Sitter brane. We shall include a coupling to a brane scalar field
$\varphi$, representing the matter fields. In general, $\Psi$ has
a bound state, whose mass $m_\bs$ depends on the vacuum
expectation value (vev) of $\varphi$. We compute the 1-loop
effective potential $V_\eff(\varphi)$ that the KK modes induce on
$\varphi$. The main result that we find is that $V_\eff(\varphi)$
develops a 'true' vacuum with $\la\varphi\ra\neq0$, rendering the
original vacuum $\la\varphi\ra=0$ meta-stable. In the new vacuum,
the bulk field $\Psi$ does not have a light mode anymore.

Intuitively, the meta-stability of the $\varphi=0$ vacuum can be
understood as follows. In our model, we take an interaction on the
brane of the form
$$
\lambda\;\varphi^2 \,\Psi^2|_0~~,
$$
where $|_0$ means that the bulk field is evaluated on the bane.
The effect of the KK modes of $\Psi$ can be accounted for
substituting $\Psi^2|_0$ by its vacuum expectation value
$\phisq|_0$. This effectively acts as a ($\varphi-$ dependent)
mass term for $\varphi$. For $\varphi\to0$ the bound state of
$\Psi$ is light, $m_\bs\ll H$, and $\phisq|_0$ is large. Hence,
$m_\varphi^2(\varphi)$ is decreases with $\varphi$, close to
$\varphi=0$. For large $\varphi$, $\phisq|_0$ grows (as it does in
flat space). Hence, $m_\varphi^2(\varphi)$ has a minimum. If this
is deep enough then $\varphi$ is naturally driven towards it.
Hence, at 1-loop level, the $\varphi=0$ vacuum is not stable, and
a true vacuum must appear for some $\varphi\neq0$. Note that it is
important that the brane geometry is close to de-Sitter, otherwise
the fluctuations of light modes are not large. Applications to
cosmological models are discussed in the conclusions.\\

Previous works on quantum effects in the BW scenario have focused
on the Casimir force experienced by the branes in RSI-type models
\cite{gpt,flachitoms,gr} and generalizations of this setup (see
{\em e.g.} \cite{gpt2,fgpt,fp,gapo,brevik,sase}). Explicit results
for cosmological branes are available only for de Sitter branes in
AdS${}_5$ with conformally coupled bulk fields
\cite{wade,nojiri,elizalde,ns,fkns}, or for generic fields in a
flat bulk \cite{pt,xavi}. In particular, in the one brane case the
fluctuations on the brane $\phisq|_0$ were found to vanish for a
conformally coupled bulk field \cite{ns}. This should be
interpreted to mean that the contribution is purely local and it
can be 'renormalized away' by a finite renormalization of (local)
counter-terms. In this article, we consider a conformal field that
is non-trivially coupled to the brane with a brane-mass $m$. Then
$\phisq$ includes a non-local part that can be related to the
effective potential.

We shall drive the attention to a technical remark. The
computation of the quantum fluctuations on the brane $\phisq|_0$
faces a subtlety in the \emph{thin wall} approximation. In this
treatment, the effect of the branes is treated by boundary
conditions where the field (and/or its derivative) is forced to
obey some condition at the brane location. A generic consequence
of this is that the vev of the field fluctuations $\phisq$ (as
well as $\tmn$) blow up close to the brane \cite{bida,dc,kcd} (see
also \cite{romeosaharian,fulling}). This has been recently
manifested in the context of the RSI model in \cite{knapman} and
is of concern because it is precisely the quantities evaluated on
the brane that are directly coupled to matter in the Brane world
scenario. One way to compute $\phisq|_0$ (evaluated on the brane)
is to take into account the brane thickness \cite{olum}. In this
article, we shall see that in order to regularize $\phisq|_0$ in
the thin wall approximation, one needs to subtract divergences
given by the extrinsic curvature of the brane as well as the mass
of the field. Accordingly, $\phisq|_0$ is well defined up to
finite renormalization of mass and extrinsic curvature terms,
and still we can extract some physical information.\\

This article is organized as follows. In Section \ref{sec:fluct},
we compute the fluctuations of the scalar field in the bulk
$\phisq(z)$ and on the brane $\phisq|_0$, using a number of
regularization schemes and we unveil the connection between them.
For the bulk field, we consider conformal coupling with zero bulk
mass but non-vanishing brane mass because this case admits an
analytic treatment. In Section \ref{sec:model}, we describe the
interacting model, and discuss the 4D effective theory obtained by
the Kaluza-Klein reduction (also called dimensional reduction) and
by the geometrical projection method
\cite{sms,maedawands,ls,soda}. In Section \ref{sec:veff}, we
present the results for the effective potential $V_\eff(\varphi)$,
and we conclude with some remarks and applications in Section
\ref{sec:concl}.

\section{Quantum fluctuations from a bulk field $\Psi$}
\label{sec:fluct}

We consider a scalar field propagating in the bulk described  by
\begin{equation}\label{actionPhi}
    S_{\Psi}=-{1\over2}\int d^{n+2}x \;\sqrt{-g}
    \left[ \lp(\partial\Psi\rp)^2+
   \left(  M^2 +\xi R\right) \Psi^2 \right]
    -\int d^{n+1}x \;\sqrt{-h} \;
\left[\mbr+2\xi \, K \right]
        \,\Psi^2~,
\end{equation}
where $M$ is the bulk mass, $\xi$ is the nonminimal coupling, $K$
denotes the trace of the extrinsic curvature $K_{\mu\nu}$ and we
allow for a brane mass $\mbr$. Here, $g$ and $h$ denote the
determinants of the metric on the bulk $g_{\mu\nu}$ and of the
induced metric on the brane $h_{\mu\nu}$. For $AdS_{(n+2)}$ bulk,
the line element is
\begin{equation}
\label{metric}%
ds^2
=a^2(z)[dz^2+ds^2_{(n+1)}]
\end{equation}
where $a(z)= 
\ell/ \sinh (z_0+z)$, $\sinh z_0= H\ell$, $\ell$ is the AdS radius
and $H$ is the Hubble constant on the brane. We denote by
$ds^2_{(n+1)}$ the metric on an $(n+1)-$dimensional de Sitter
space of unit radius, and $z=0$ is the brane location. The
Klein-Gordon equation is separable, so the field admits a mode
decomposition of the form $\Psi(z,x^\mu)=\sum_p
\uu_p(z)\Psi_p(x^\mu)$ where the sum runs over the spectrum.
The $(n+1)-$dimensional modes are labelled by $p$ and have masses
$m_\kk^2=[(n/2)^2+p^2]H^2$. The mass spectrum is determined by the
radial equation together with the boundary conditions. For
simplicity, from now on we shall concentrate on the case of
conformal coupling
$$
\xi_c={n\over4(n+1)}~,
$$
and vanishing bulk mass $M$, though we will allow for a nonzero
brane mass. In this case, the boundary condition on the brane is
\beq%
\label{bc}%
\lp[\partial_z+\nu \rp]\big(a^{n/2} \uu_p \big) \big|_0=0
\eeq%
with
\beq%
\label{nu}%
\nu\equiv -{\mbr \over H}
\eeq%
and in (\ref{bc}), $|_0$ denotes the quantities that are evaluated
on the brane. %
The radial dependence of the KK modes takes the simple form
\begin{equation}
\label{kkwf}
 \uu^\kk_p(z)=\sqrt{a^{-n}\over \pi
   (1+(\mmu/p)^2)}\left(
   \cos \left( p z  \right)
         -{\mmu\over p}\sin \left( p z   \right)\right)~,
\end{equation}
with $p>0$.
When $\nu>0$, a normalizable 'bound state' exists at $p=i\nu$. Its
mass is of the form
\beq\label{mbs}%
m_\bs^2=\lp[(n/2)^2-\nu^2\rp]H^2~,%
\eeq%
and its wavefunction is
\beq\label{bswf}%
\uu^{\bs}= \sqrt{\nu} a^{-n/2} e^{-\nu z}~.
\eeq%
For $\nu<0$, the bound state disappears from the spectrum because
it becomes un-normalizable. The mode with $p=i\nu$ (in the lower
half $p-$plane) corresponds to a quasi-normal mode
\cite{rubakov,ls}, and satisfies the purely outgoing-wave boundary
condition at the future Cauchy horizon of $AdS_5$. Note that in
our case, this mode does not have oscillatory part, so it is
'purely decaying'. Thus, even though $p$ is pure imaginary, the
mass squared of this quasi-normal mode takes the same form as
(\ref{mbs}).

\subsection{$\phisq$ in the bulk}

In the mode sum representation, we must include the contributions
from the bound state (if any) and the KK modes,
\begin{equation}\label{greg}
\begin{array}{rcll}
  G_{}^{(1)}&=&G^{\kk}+G^{bs}~, & \textrm{with} \\[2mm]
  G^{bs}&=&\uu^\bs(z)\uu^\bs(z')G_{i\nu\,(dS)}^{(1)}&\textrm{and} \\[2mm]
  G^{\kk}&=&\displaystyle\int_0^\infty dp \;\uu^\kk_p(z )
      \uu^\kk_p(z')\;G_{p\,(dS)}^{(1)}~,&
\end{array}
\end{equation}
where $G_{p\,(dS)}^{(1)}$ denotes the Bunch-Davies Green function
for the corresponding mode. It is understood that $G^\bs$ should
be included only for $\nu>0$. One easily finds \cite{pt}
$$
\uu^\kk_p(z) \uu^\kk_p(z')=\lp[\uu^\kk_p(z)
\uu^\kk_p(z')\rp]_++(p\to-p)~,
$$
with
\begin{equation}
\label{greg0}
\lp[\uu^\kk_p(z) \uu^\kk_p(z')\rp]_+%
   ={1\over 4\pi (aa')^{n/2} }
    \left[{ p +i\nu\over p-i\mmu}e^{i(z+z')p}+e^{i(z-z')p}\right]~.
\end{equation}
and we can write $G^{\kk}=\int_{-\infty}^\infty dp \lp[\uu^\kk_p(z
) \uu^\kk_p(z')\rp]_+G_{p\,(dS)}^{(1)}$. In order to compute the
regularized Green function, we have to subtract the divergence
present in the absence of the brane. In that case, the modes look
like $\uu^{0}_p=e^{ipz}/\sqrt{4\pi a^n}$ with either positive and
negative $p$, and are normalized according to
$2\int_{-\infty}^{\infty}dz a^n
\uu^{0}_p\uu^{0}_{p'}{}^*=\delta(p-p')$. Thus, the regularized
Green function is%
\beq%
\label{gren}%
G_{reg}^{(1)}=G^\bs+\int_{-\infty}^\infty dp \lp[\uu^\kk_p(z)
\uu^\kk_p(z')\rp]^{reg}G_{p\,(dS)}^{(1)} %
\eeq%
where
\begin{equation}
\lp[\uu^\kk_p(z) \uu^\kk_p(z')\rp]^{reg}%
\equiv \lp[\uu^\kk_p(z) \uu^\kk_p(z')\rp]_+-\uu^0_p(z)
{\uu^0_p}^*(z')
   ={1\over 4\pi (aa')^{n/2} }
    { p +i\nu\over p-i\mmu}e^{i(z+z')p}~.
\end{equation}
One can now perform the $p$ integral in (\ref{greg}) closing the
contour in the complex $p$ plane by the upper half plane. One then
sums residues of the poles. The residue from the pole due to
$\lp[\uu^\kk_p(r) \uu^\kk_p(r')\rp]^{reg}$ at $p=i\nu$ turns out
to cancel the contribution from the bound state, and we have to
sum over the poles arising from $G_{p\,(dS)}^{(1)}$ only. The
massive Wightman function in dS space is\footnote{Note that there
is a typo in Eq. (A5) of \cite{pt}-- a factor 2 is missing in the
rhs.} \cite{pt}
\begin{align}
\label{gds}
 G^{(dS)(1)}_p(x,x')&={2\over (n-1)S_{(n)}}%
 {2^{-{n-1\over2}}\over(1-\cos\zeta)^{\q-1\over 2}}
      \Biggl\{F\left(-ip+{1\over 2},ip+{1\over 2},{-\q+3\over 2};
     {1-\cos\zeta\over 2}\right)\cr %
     &+
     {\Gamma\left({\q\over 2}-ip\right)\Gamma\left({\q\over 2}+ip\right)
          \Gamma\left(-{\q-1\over 2}\right)
     \over \Gamma\left({1\over 2}-ip\right)\Gamma\left({1\over 2}+ip\right)
          \Gamma\left({\q-1\over 2}\right)
      }\left({1-\cos\zeta\over 2}\right)^{\q-1\over 2}
      F\left(-ip+{\q\over 2},ip+{\q\over 2},{\q+1\over 2};
     {1-\cos\zeta\over 2}\right)
   \Biggr\}
\end{align}
where $F$ is the hypergeometric function and $\zeta$ is the
invariant distance between $x$ and $x'$. Using (\ref{gds}), we
obtain \cite{pt}
\begin{align}
\label{bulk}
  G_{reg}^{(1)}
   =  &{ 1\over   S_{(\q)}\Gamma\left({n+1\over
   2}\right)}{\lp(e^{-(z+z')}\over 4a a'\rp)^{n/2}}
          \sum_{\s=0}^{\infty}    \sum_{j=0}^{\infty}
          {{\q\over 2}+\mmu+\s+j \over {\q\over 2}-\mmu+\s+j}
          ~{(-1)^\s \Gamma\left(\q+2\s+j\right)
            \over j!\,\s !\, \Gamma\left({n+1\over 2}+\s\right)}
          e^{-(\s+j)(z+z')}
          \left({1-\cos\zeta\over 2}\right)^\s~.
\end{align}
This can be summed for any value of $n$ because $e^{-(z+z')}<1$.
The result for $\nu=0$ is
\begin{align}
\label{Gconf}
  G^{(1)}_{\nu=0}(x,x')&=
{1\over n S_{(n+1)}}   \lp( {1\over \lp( a a' e^{z+z'} \rp) \lp(
1+ e^{-2(z+z')} - 2 e^{-z-z'} \cos\zeta \rp)}\rp)^{{n / 2}},
\end{align}
where $\zeta$ is the invariant distance in $dS$ space and
$S_{(n+1)}=2 \pi^{1+n/2}/\Gamma(1+n/2)$ is the volume of a unit
$n+1$ dimensional sphere. Equation (\ref{Gconf}) can be easily
derived by the method of images and making a conformal
transformation to flat space (see Appendix (\ref{app:conf})).

We see that the coincidence limit $z=z'$, $\zeta=0$ is finite as
long as we are not on the brane, at $z=0$. This readily provides
the result for the fluctuation of the field in the bulk as
\begin{align}
\label{coinc}
 \phisq(z)&= {1\over2} G_{(ren)}^{(1)}(z,z,0)
   =  { 1\over   2S_{(\q)}\Gamma\left({n+1\over
   2}\right)^2}
{\lp(e^{-z}\over 2a \rp)^{n}}
          \sum_{j=0}^{\infty}
          {{\q\over 2}+\mmu+j \over {\q\over 2}-\mmu+j}
          ~{ \Gamma\left(\q+j\right)
            \over j!}
          \;e^{-2j z}\cr
&= { 1\over  2n S_{(\q+1)}} {1\over \lp(e^{z}a \rp)^{n}} \lp[
{1\over
(1-e^{-2z})^{n}}+{4\nu\over n-2\nu} %
\; F\lp(n,n/2-\nu,1+n/2-\nu;e^{-2z}\rp)\rp] ~.
\end{align}
For $\nu=0$, corresponding to a truly conformally coupled field,
one has
\begin{align}
\label{phi2flat} %
\phisq(z)
={1\over 2n S_{(n+1)} \,%
\left[2 a(z) \sinh(z)  \right]^{n}}~.
\end{align}
Note that this result as well as (\ref{coinc}) are independent of
the form of the warp factor $a(z)$. It holds for any space of the
form (\ref{metric}).

Equation (\ref{phi2flat}) agrees with \cite{ns}, where $\phisq$ is
computed using a different method. There, a regulating brane is
introduced, and the computation is made in a conformally related
space. Then, the regulating brane is sent to infinity. This method
was shown to reproduce incorrect results for global quantities
such as the effective action \cite{fkns}. The reason is that this
procedure does not preserve the topology, because the conformal
transformation at infinity is divergent. However, the agreement in
the computation of $\phisq(z)$ suggests that the procedure based
on the regulating brane still works to compute local quantities.
In Appendix \ref{app:conf}, we re-derive the same result using a
conformal transformation that it is regular on all the points of
the manifold, which guarantees that the topology is preserved.

\subsection{$\phisq$ on the brane}

Our aim is to find the value of $\phisq$ when the field is
restricted on the brane. The limit $z\to0$ of (\ref{coinc})
diverges. In order to compute the renormalized value of the
fluctuations on the brane $\phisq|_0$ we need to perform some
further subtraction. One possibility is to remove from
(\ref{coinc}) the terms that diverge in the limit $z\to0$. Another
possibility is to restrict from the beginning to the fluctuations
on the brane and do the computation using dimensionally
regularization. In this Subsection, we shall pursue both schemes,
and show that they agree up to finite renormalization of local
counter-terms.

We can obtain the fluctuations restricted on the brane essentially
as in the previous Subsection. From the form of the KK modes, we
have
\begin{equation}
\label{uu}
 \lp[\uu^\kk_p \uu^\kk_p{}'\rp]^{reg}\Big|_{0}
={H^{{n}}\over 4\pi  }
    { p +i\nu\over p-i\mmu}~.
\end{equation}
Proceeding as in (\ref{greg}) and performing the $p$ integration,
we readily obtain for the Green function restricted on the brane,
\begin{align}
\label{brane}
  &G_{}^{(1)}\big(z=z'=0,\zeta\big)
   =  { (H/2)^{n}\over   S_{(\q)}\Gamma\left({n+1\over
   2}\right)}
          \sum_{\s=0}^{\infty}    \sum_{j=0}^{\infty}
          {{\q\over 2}+\nu+\s+j \over {\q\over 2}-\mmu+\s+j}
          ~{(-1)^\s \Gamma\left(\q+2\s+j\right)
            \over j!\,\s !\, \Gamma\left({n+1\over 2}+\s\right)}
          \left({1-\cos\zeta\over 2}\right)^\s
\end{align}
In contrast with (\ref{bulk}), the sum over $j$ now diverges.
There are several ways to regularize this expression. One
possibility is dimensional regularization. One regards $n$ as a
complex number different from 3. For small enough values, the $j$
sum converges, the coincidence limit $\zeta\to0$ can be taken,
\begin{align}
  G_{}^{(1)}\big(z=z'=0,\zeta=0\big)
   &=  { H^{n}\over   n S_{(\q+1)}\Gamma(n)} \sum_{j=0}^{\infty}
          {{\q\over 2}+\nu +j \over {\q\over 2}-\mmu+ j}
          ~{ \Gamma\left(\q +j\right)
            \over j! }
\end{align}
and the result is analytically continued to $n=3$. We obtain
\footnote{In our case, we could have avoided the steps leading to
(\ref{general}) because the expression (\ref{coinc}) reproduces
this result quite directly. The limit $z\to0$ appears divergent,
but the dimensionally regularized value is well defined. For
negative values of $n$, the first term in (\ref{coinc}) vanishes,
and the second coincides with Eq.
(\ref{general}) in this limit.}%
\begin{align}
\label{general}%
\phisq\big|_{0} %
%
={ H^{n}\over n S_{(n+1)}}%
\;{\nu\Gamma(n/2-\nu)\Gamma(1-n)\over %
\Gamma\lp(1-n/2-\nu\rp)}
\end{align}

Expanding this expression around $n=3$, we find
\begin{align}
\phisq\big|_{0}&= \left(H/\mu\right)^{n-3} { H^{3}\over
8 \pi^2}%
\Big\{ {\nu-4\nu^3\over 8}{1\over n-3}
+{\nu-4\nu^3\over8} \left[ \psi(3/2-\nu)+c\right] -
{\nu^2\over2}+{\cal O}\lp(n-3\rp) \Big\}\nonumber
\end{align}
where $\psi(z)=\Gamma'(z)/\Gamma(z)$ is the digamma function, $c$
is an irrelevant numerical constant. We have introduced an
arbitrary renormalization scale $\mu$ in order that $\phisq$ keeps
the dimensions of mass cubed. Dropping the pole in $n-3$, we
obtain on the brane
\begin{equation}
\label{resultdimreg}%
{}^{(ren)}\phisq\big|_{0}= {  -4\nu^2+(\nu-4\nu^3) \left[
\psi(3/2-\nu)+\ln(H/\mu)\right] \over
64 \pi^2 H^{-3}}~,%
\end{equation}
where we made a finite and constant redefinition of $\mu$.

The renormalization procedure becomes clearer using a cutoff
scheme. A cutoff in $p$ is equivalent to introduce a cutoff in the
summation (\ref{brane}), the latter being more convenient. Setting
$n=3$ and $\zeta=0$, we obtain
\begin{align}\label{cutoff}
      \phisq\big|_0
    =  {H^{3}\over   32\pi^2} &\sum_{j=0}^{J} {{3\over 2}+\nu+j \over {3\over 2}-\mmu+j}
           ~(j+1)(j+2)\cr
    = {H^{3}\over   64\pi^2} &\Biggl\{ {2J^3/3}+ 2(\nu+2) J^2+2\big(2\nu^2+4\nu+11/3\big) J \cr
    &+  4+ 6\nu+4\nu^2+(\nu-4\nu^3) \left[
\psi(3/2-\nu)-\ln{J}\right]  + {\cal O}\lp(1/J\rp) \Biggr\}~.%
\end{align}
In the context of the scalar model of Section \ref{sec:model}, the
linear, quadratic and cubic divergences in $\nu$ can be cancelled
by appropriate counter-terms in the Lagrangian. In turn, this
means that the actual value of the constant, $\nu$ and $\nu^2$
terms are not really physical and have to be fixed by
renormalization conditions.

There exists still another regularization procedure, which
consists in taking the limit of the bulk contributions $\phisq(z)$
for $z\to0$,
\begin{align}\label{pointsplit}
\phisq(z)&\sim {H^3\over 128 \pi^2}\Biggl[{1\over z^3}+ {3 \yy -2
\nu\over z^2}+ {1+ 3\yy^2+6\yy\nu+4\nu^2\over z}\cr%
&+\yy^3+6\yy^2\nu+2\yy(1+6\nu^2)%
-{2\over3}\nu (1+12\nu)%
+ 2\nu(1-4\nu^2) \left[ \psi(3/2-\nu)+\ln{2z}+\gamma\right] %
+{\cal O}(z)\Biggr]~,
\end{align}
where %
\beq\label{y}%
y\equiv {K\over 4H}=\sqrt{1+(H\ell)^{-2}} ~.
\eeq%

We can see the agreement in the finite part up to finite
renormalization of mass and extrinsic curvature terms, and it
explicitly shows that in principle the extrinsic curvature terms
appear in $\phisq|_0$.
This suggests that in order to compute the quantities restricted
on the brane, it is enough to compute the corresponding quantity
in the bulk, and remove the terms that diverge close to the brane.
This seems reasonable since, after all, one can regard this
procedure as a sort of point splitting regularization. More
technically, this happens in our case because the Green function
in the bulk (\ref{bulk}) is equivalent to the Green function
restricted to the brane with an exponential suppression of each
term, which one expects that should be equivalent to introducing a
cutoff. It is likely that this equivalence holds in situations
with less symmetry.

~\\

We shall conclude this Subsection by writing down the form of
$\phisq|_0$ derived from (\ref{resultdimreg}), (\ref{cutoff}) and
(\ref{pointsplit}):
\begin{equation}
\label{result2}%
{}^{(ren)}\phisq\big|_{0}= {  \nu(1-4\nu^2) \left[
\psi({3\over2}-\nu)+\ln(H/\mu)\right] + A+ B\nu+ C\nu^2\over
64 \pi^2 }\;H^{3}~,%
\end{equation}
where $A$, $B$, $C$ and $\mu$ have to be fixed by renormalization
conditions and $\nu$ is given by (\ref{nu}). We shall return to
this issue in Section \ref{sec:model}. In Appendix \ref{app:conf},
we comment on a check of (\ref{result2}) based on a conformal
transformation and previous results in the literature.

\subsection{Bound state and KK contributions}
\label{sec:bskk}

To obtain the contribution from the bound state
${}^{(\bs)}\phisq|_0$, we just multiply the fluctuations from this
4D mode by the wave-function squared, $\uu_\bs^2|_0=\nu H^3$. From
the well known form of the fluctuations of a scalar field in 4D de
Sitter space \cite{bd,vilenkinford}, we find
\begin{align}
\label{bsalone} %
{}^{(\bs)}\phisq|_0&= \theta(\nu)\,{\nu \left(1-4\nu^2\right)
\left[
\psi\left({3\over2}+\nu\right)+\psi\left({3\over2}-\nu\right)%
+2\ln\left(H/\mu\right)%
 \right]\over 64\pi^2 }\,H^{3} ~,
\end{align}
up to finite renormalization of local terms. The step function
$\theta(\nu)$ ensures that for there is no contribution for
$\nu<0$. Comparing with (\ref{result2}), and splitting
$\phisq|_0={}^{(\bs)}\phisq|_0+{}^{(\kk)}\phisq|_0$, we identify
the KK contribution as
\beq%
\label{kk}%
{}^{(\kk)}\phisq\big|_{0}= {  -|\nu|(1-4\nu^2) \left[
\psi({3\over2}+|\nu|)+ \ln(H/\mu)\right] \over
64 \pi^2 H^{-3}}~.%
\eeq%
We can recognize in (\ref{bsalone}) the contributions from the
growing and decaying modes (of the homogeneous mode) as the
$\psi(3/2\mp\nu)$ respectively. Thus, when the bound state exists
($\nu>0$), the contribution from all the KK modes equals to minus
the decaying mode of the bound state. For $\nu<0$, the KK modes
behave like a decaying mode with mass-squared
$\lp[(3/2)^2-\nu^2\rp]H^2$. This mode is naturally identified as
the quasi-normal mode \cite{rubakov,ls} mentioned above, which is
consistent with the fact that this mode is purely decaying in our
case.

\begin{figure}[htb]
  \includegraphics[width=7cm]{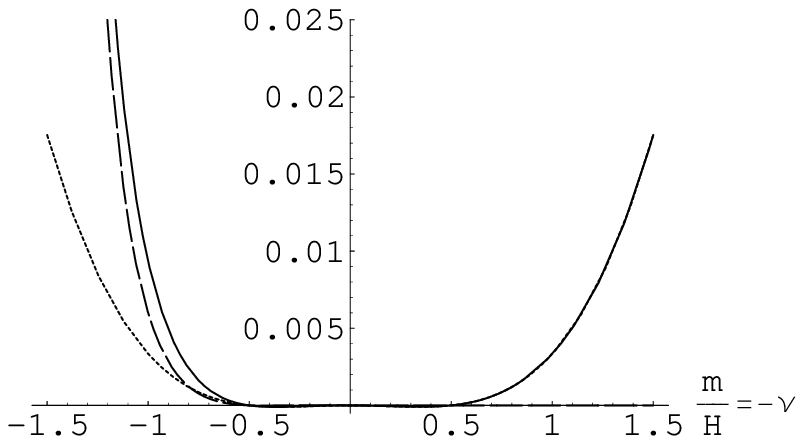}\\
  \caption{
   The solid line is $\phisq|_0$, the dashed line
   is the bound state contribution ${}^{(\bs)}\phisq|_0$ (which is absent for
   $m>0$), and the dotted line corresponds to the KK contribution
   ${}^{(\kk)}\phisq|_0$.
   These values are for $\mu=H$ and are expressed in units of $H^3$.
   For $\mbr\to-3H/2$, the mass of the bound state
    vanishes like $m_\bs^2\simeq 3(m+3H/2)H$, and $\phisq|_0$ and
    ${}^{(\bs)}\phisq|_0$ grow like $\sim m_\bs^{-2}$.
    In the same limit, the KK contribution grows but remains finite.
   }\label{fig:phi2}
\end{figure}

Figure \ref{fig:phi2} shows that in the limit $m_\bs\to0$
($\nu\to3/2$), the contribution from the KK modes
(\ref{kk}) is negligible relative to the bound state
contribution 
(\ref{bsalone}). In this limit, the bound state contribution
diverges like ${}^{\bs}\phisq|_0\sim (9/16\pi^2) (H^5/m_\bs^2)$,
whereas the KK contribution grows but stays of order
${}^{\kk}\phisq|_0\sim (27/128\pi^2)H^3$. Note that this statement
does not depend on the choice of renormalization coefficients in
(\ref{result2}) (also present in (\ref{bsalone}) and (\ref{kk})).
This result agrees with \cite{kks,hts}, where the contribution
from the KK modes to the power spectrum of primordial fluctuations
in the bulk inflaton model \cite{hs,kks} was also found to be
small when the bound state is light.

\subsubsection*{Four dimensional conformal coupling}

We see that the divergent part of (\ref{general}) vanishes for
$\nu=0,\pm1/2$. In these cases, $\phisq|_{}$ does not depend on
the renormalization scale $\mu$, and one can try to assign to
$\phisq|_{}$ an unambiguous value. If we take the result from
dimensional regularization (\ref{general}), then $\phisq|_{}=0$
for $\nu=0$. In this case, the bound state is not normalizable
($\uu_\bs=0$), and we conclude that the KK contribution also
vanishes. In light of the relation between dimensional
regularization and other schemes, this means that $\phisq|_0$ in
this case is a pure counter-term.

For $\nu=\pm1/2$, Eq (\ref{general}) gives
\begin{equation}
\label{nu12}%
\phisq\big|_{0}^{(\nu=\pm1/2)}= -{ H^{3}\over
64 \pi^2}~.%
\end{equation}
For $\nu=-1/2$, there is no bound state, and the above is the
contribution from the KK modes. The case $\nu=1/2$ corresponds to
a bound state that effectively is conformally coupled in the four
dimensional sense because $m^2_{\bs}=2H^2=R_{(4)}/6$ (in 4D
conformal coupling is for $\xi=1/6$). It is fortunate that in this
case the five dimensional result is 'finite' (there is no $\mu$
dependence) because the fluctuations of a conformally coupled
scalar in dS are also finite \cite{vilenkinford}, allowing for a
straightforward comparison. The fluctuations for a conformally
coupled scalar in four dimensional dS of unit radius give
$1/48\pi^2$ \cite{vilenkinford}. In our model, this contribution
to the fluctuation has to be weighted by the wave function of the
mode at the brane location $\uu_\bs^2|_{0}=\nu H^{n}$, so the
bound state contributes as $ {}^{(\bs)}\phisq|_{0} =H^3/96\pi^2$.
The contribution from the KK modes then is $ {}^{(kk)}\phisq|_{0}
=-(5/2)\;{}^{(\bs)}\phisq|_{0}$. Thus, the 'correction' from the
KK modes in this case is rather large. This could be anticipated
because in this case the bound state is not very light,
$m_\bs=\sqrt{2}H$, so we do not expect it to dominate over the KK
modes.

It is also worth noting that (\ref{nu12}) is negative (because the
KK contribution is negative). This is a typical outcome of the
procedure of regularization necessary to make sense of divergent
sums, even when all the terms are positive definite. So it should
be interpreted with care. For instance, we can always choose a set
of renormalization conditions (or make a finite renormalization of
local counter-terms) so that $\phisq$ becomes positive.

\section{Bulk--brane interaction and the effective potential}
\label{sec:model}

We shall consider a bi-scalar model
\beq%
\label{model}%
S=S_\Psi+S_\varphi+S_{int}%
\eeq%
where $S_\Psi$ is given in (\ref{actionPhi}), and for the brane
field $\varphi$
\begin{equation}\label{actionPsi}
    S_{\varphi}=-{1\over2}\int d^4x\sqrt{-h}
    \left[(\partial \varphi)^2 +
m_\varphi^2\varphi^2\right]~,
\end{equation}
where $h$ denotes the determinant of the induced metric on the
brane.
As for the interaction term, we take%
\beq
\label{int}%
 S_{int}=-\int d^4x\sqrt{-h}\; \lambda\, \varphi^2
\,\Psi^2|_0%
\eeq%
where $\Psi^2|_0$ stands for the bulk field evaluated on the
brane, and
$\lambda$ is a coupling constant with dimensions of length. %

\subsection{Kaluzaz-Klein Reduction}

The usual 'Kaluza-Klein decomposition'  (also called dimensional
reduction) consists in inserting the KK ansatz
$\Psi(z,x^\mu)=\sum_p \uu_p(z)\Psi_p(x^\mu)$ with $\uu_p(z)$ given
by (\ref{bswf}) and (\ref{kkwf}), introduce it into the action
(\ref{model}) and integrate out the extra dimension. Because of
the orthonormality of the wavefunctions $\uu_p(z)$,
the resulting 4D action at quadratic order is%
\beq\label{sDR} %
S=-{1\over2}\int \sqrt{h}d^4x\;%
\lp\{ \lp(\partial\varphi\rp)^2 +m_\varphi^2 \varphi^2+
\lp(\partial\Psi_\bs\rp)^2 +m_\bs^2 \Psi_\bs^2~+~ {\rm KK~modes}
\rp\}
\eeq %
where $m_\bs^2=(3H/2)^2-\mbr^2$. Having a light bound state
$m_\bs\ll H$, reduces to choosing $\mbr$ close enough to $-3H/2$.
One expects that this models some of the features of the minimally
coupled case.

Restricting ourselves to configurations of constant $\varphi$, the
interaction (\ref{int}) can be taken into account by the
replacement
\begin{align}\label{repl}%
\mbr&\to\mbr+\lambda\varphi^2~.
\end{align}
Then, the mass of the bound state is given by
\beq\label{spectrum}%
m_\bs^2=(3H/2)^2-\lp(\mbr+ \lambda \varphi^2 \rp)^2~.%
\eeq%
and we can identify the \emph{classical} potential as
\beq%
\label{vDR}%
V^{cl}={1\over2}m_\bs^2 \Psi_\bs^2 +{1\over2}m_\varphi^2 \varphi^2
={1\over2}m_{0}^2\Psi_\bs^2
+{1\over2}m_\varphi^2 \varphi^2%
- {\mbr \lambda}\,\varphi^2 \Psi_\bs^2 %
- {1\over2} \lambda^2 \varphi^4 \Psi_\bs^2 %
\eeq%
where $m_{0}^2=\lp(3H/2\rp)^2-\mbr^2$.  Note that this potential
contains a biquadratic interaction similar to the one in
(\ref{int}) with an effective (dimensionless) coupling constant
given by $-m \lambda$ (recall that $\mbr<0$). Aside from it, we
notice an extra piece $\propto -\lambda^2 \varphi^4 \Psi_\bs^2$.
This term can be interpreted as a higher dimensional effect.
%
To see this, note that it is crucial wether we consider the
interaction (\ref{int}) 'turned on' at the 5D level (before doing
the dimensional reduction), or at the 4D level (once it is already
done). If it is considered turned off when doing the reduction,
then $m_\bs^2=(3H/2)^2-\mbr^2$ and $\uu_\bs^2|_0\propto- \mbr $
(see Eqns. (\ref{bswf}) and (\ref{nu})). To include the
interaction, we insert this decomposition in (\ref{int}) and the
only interaction with the bound state $\Psi_\bs$ is $\lambda
\,\uu_\bs^2|_0 \,\Psi_\bs^2\,\varphi^2$, which agrees with the
third term in (\ref{vDR}). Thus, in the '4D treatment', the
interaction is the bi-quadratic term only. If we consider
(\ref{int}) turned on at the 5D level, the spectrum
(\ref{spectrum}) depends on $\lambda\varphi^2$. In particular
$\uu_\bs^2|_0\propto- \mbr - \lambda\varphi^2$, whence the new
term arises. No correction is obtained unless the interaction is
considered in the 5D sense, so we interpret the last term in
(\ref{vDR}) as a higher dimensional effect.

From the AdS/CFT correspondence \cite{adscft}, the $\lambda^2
\Psi_\bs^2 \varphi^4$ term can be interpreted as a quantum
correction from the CFT, which agrees with the fact that it is of
order $\lambda^2$. We leave for future investigation a detailed
analysis of the correspondence in this setup. In Section
\ref{sec:geom}, we give further evidence for this interpretation,
showing that one can reproduce the above potential using the
method based on the geometrical projection of \cite{sms}, as done
in \cite{maedawands,ls}. In this approach, these terms arise from
the square of the matter stress tensor (which are related to the
conformal anomaly \cite{shiromizu}) and from the normal
derivatives of $\Psi$ present in the bulk stress tensor
\cite{maedawands}. The analysis made in \cite{ls} reveals that the
effective potential obtained in this way agrees with that obtained
by the mode decomposition, at least to leading order in the
coupling to the brane ($m$, in our model).

Finally, note that the potential (\ref{vDR}) is unbounded from
below. This is not a problem, for two reasons. As we show in
Section \ref{sec:geom}, when we take into account all the terms in
the effective potential as derived with the geometrical projection
method \cite{maedawands,ls}, then it becomes bounded. Moreover,
when the unbounded term becomes noticeable
$(\lambda\varphi^2\gtrsim H )$ the bound state mass is comparable
to $H$ and eventually disappears as a normalizable mode, meaning
that the effective description (\ref{sDR}) and (\ref{vDR}) breaks
down.

\subsection{Geometrical Projection method}
\label{sec:geom}

Here, we discuss how the previous effective theory can be
partially re-derived by means of the geometrical projection of the
equations of motion on the brane \cite{sms}. The dilaton-gravity
system in the BW was studied in \cite{maedawands}, and the form of
the effective potential for the four dimensional dilaton field was
derived. In \cite{ls} (see also \cite{shs}), this was compared to
the mode spectrum, and the two approaches were found to agree at
the linear level in the brane coupling ($m$ in our notation). In
\cite{soda}, the connection between this approach and the gradient
expansion method \cite{kannosoda} is described. References
\cite{maedawands,ls,soda} considered a minimally coupled field in
the bulk. In this section we extend their analysis to the
conformally coupled case.

In \cite{sms}, the effective 4D Einstein equations were found to
be
\begin{equation}
{}^{(4)}G_{\mu\nu}=\kappa_5^2
T^{Proj}_{\mu\nu} 
+ 8 \pi G_N\;T^{Brane}_{\mu\nu}+\kappa_5^4\,\pi_{\mu\nu}
-E_{\mu\nu}
\end{equation}
where $T^{Brane}_{\mu\nu}$ is the matter stress tensor on the
brane, $\pi_{\mu\nu}$ is quadratic in $T^{brane}_{\mu\nu}$,
$E_{\mu\nu}$ is the projected Weyl tensor, and
\begin{equation}
T^{Proj}_{\mu\nu}\equiv
\frac{2}{3}\lp\{T^{Bulk}_{\rho\sigma}h_{~\mu}^\rho
h_{~\nu}^\sigma+ \big( T^{Bulk}_{\rho\sigma}n^{\rho}n^{\sigma}
-{1\over 4}T^{Bulk\,\rho}_{~\rho} \big) h_{\mu\nu} \rp\}~.
\end{equation}
Here, $n^\mu$ is unit vector normal to the brane and
$T^{Bulk}_{\rho\sigma}$ is the bulk stress tensor, and the
effective Newton's constant is $G_N=\kappa_5^4\,\sigma/48 \pi$
where $\sigma$ is the brane tension.

The contribution to the stress tensor from a nonminimally coupled
bulk field can be concisely written as \cite{bida,saharian}.
\begin{eqnarray}\label{tmnbulk}
T_{\mu\nu}^{Bulk}&=&\partial_\mu\Psi\partial_\nu\Psi-{1\over2}\;
\left(\partial\Psi\right)^2 g_{\mu\nu}%
+\left[-{1\over2}M^2g_{\mu\nu}+\xi \lp( G_{\mu\nu}%
+ g_{\mu\nu} \Box - \nabla_\mu\nabla_\nu \rp) %
\right]\Psi^2 \\
\label{tmnbrane}%
T_{\mu\nu}^{Brane}&=&\delta(r-r_0)\lp[2\xi\,  K_{\mu\nu}
-h_{\mu\nu}\left( \mbr+2\xi K +2\xi n^\mu \partial_\mu \right)\rp]
\,\Psi^2~,
\end{eqnarray}
where $h_{\mu\nu}$ is the induced metric on the brane,
$G_{\mu\nu}$ is the bulk Einstein tensor, and $r$ denotes the
normal coordinate. See
\cite{fulling,romeosaharian,saharianWightman} and references
therein for the relevance of the surface terms in Casimir energy
computations. Note that in these expressions we didn't use the
equations of motion. For $M=\mbr=0$ and $\xi=n/4(n+1)$, and using
the equations of motion, it is easy to check that both components
are traceless. Hence, the projected bulk tensor that enters into
the effective Einstein equations is
$$
T_{\mu\nu}^{Proj}={2\over 3}\lp[h^\mu_i h^\nu_j T_{\mu\nu}^{Bulk}+
h_{\mu\nu} n^\mu n^\nu T_{\mu\nu}^{Bulk}\rp]
$$
The first two terms in (\ref{tmnbulk}) only contribute
4-dimensional derivative terms to $T^{Proj}_{\mu\nu}$.
%
%
Then, the contribution to the effective potential from $T^{Bulk}$ is%
\begin{align}\label{tproj1}%
T^{Proj}_{\mu\nu}
&={2\over 3} \xi\, h_{\mu\nu}  \lp[2 G_{rr} \Psi^2 %
+ 4 \xi R \Psi^2 +2  (\partial_r\Psi)^2  -2 \Psi \partial_r^2\Psi %
- {1\over 2} K \Psi \partial_r\Psi +\dots\rp] \cr
&= {2\over 3} \xi \,h_{\mu\nu} \lp[ -{3\over \ell^2} %
+ 2 yH \lp(m+{3\over2} yH\rp) +\dots  \rp]\Psi^2\cr
&= - {1\over2} h_{\mu\nu} \lp[ - {3\over4} H^2 %
- {1\over 2} m \,yH +\dots \rp] \Psi^2
\end{align}
where in the first equation we used the equation of motion
$\Box\Psi=\xi R\Psi$ and $K_{\mu\nu}=K h_{\mu\nu}/4$ given that
the brane is maximally symmetric. In the second, we used the
boundary condition $\partial_r \Psi|_{r=r_0^-}=-(\mbr+2\xi
K)\Psi$, Eq. (\ref{y}) and the equation of motion for the
background. For the second normal derivative we have taken
$\partial_r^2\Psi= (\mbr+2\xi K)^2\Psi$ \cite{maedawands} (see
also \cite{ls}), and the dots denote four-dimensional derivative
terms.
The brane stress tensor is
$$
T_{\mu\nu}^{Brane}%
=-\delta(r-r_0)h_{\mu\nu}\lp[\sigma+{\mbr\over 4} \,\Psi^2 \rp]
\equiv-\delta(r-r_0)h_{\mu\nu}\,\lp[\sigma+ V_0\rp]~,%
$$
where we have included the tension term. Thus, the contribution to
the effective potential from the brane
stress tensor is%
\beq\label{veffbrane} %
{\kappa_5^2\over12}\lp(\sigma+V_0 \rp)^2%
= {\kappa_5^2\over12}\lp(2\sigma V_0+V_0^2\rp) +const%
={1\over2}\lp[  {1\over 2}\mbr yH
\Psi^2+{\kappa_5^2\over6}{m^2\over16}\Psi^4+const\rp]~.%
\eeq %

From this  and Eq. (\ref{tproj1}), we find that the terms linear
in the coupling to the brane $\mbr$ in the effective mass squared
cancel. This agrees with the form of the bound state mass
(\ref{spectrum}), and also happens for the minimally coupled field
\cite{ls}. The agreement between this treatment and that of
Section \ref{sec:model} is not apparent in the higher order terms
(neither the zeroth order term, proportional to $H^2$, even though
this term vanishes in the flat brane limit). Still, the
geometrical projection method is illustrative because it unveils
the presence of interaction terms like $\Psi^4$ term in
(\ref{veffbrane}) that cannot be obtained from the mass spectrum
alone. Furthermore, the presence of $V_0^2$ in (\ref{veffbrane})
shows that the effective potential is bounded from below.

\section{1-loop effective potential}
\label{sec:veff}

The 1-loop effective potential for $\varphi$, induced by the bulk
field $\Psi$ can be obtained by the following procedure.
The equation of motion for $\varphi$ is%
\beq%
\label{eom}%
[\Box - m_\varphi^2 -2 \lambda \Psi^2|_0 ]\varphi =0.%
\eeq%
We split the brane field as $\varphi=\varphi_c+\delta\varphi$,
where $\varphi_c$ represents the vev in the true vacuum. If
$\varphi$ acquires a vev, then the effective brane mass term for
$\Psi$ is
$$
\mbr+2\xi K+\lambda\varphi_c^2~.
$$
The one loop approximation consists in replacing  $\Psi^2|_0$ by
$\phisq|_0$ in (\ref{eom}), where the latter is computed with the
effective brane mass term above. This is obtained by making the
replacement (\ref{repl})
in (\ref{result2}).
Then, from Eq. (\ref{eom}) we identify the
1-loop effective potential as
\beq%
\label{veff}%
{\partial V_\eff \over
\partial \varphi_c} \equiv m^2_\varphi \varphi_c + 2 \lambda \varphi_c\; \phisq|_0(\varphi_c)
\eeq%
Needless to say,  $\phisq|_0$ and $V_\eff(\varphi_c)$ above are
understood as renormalized values, so they depend on the
renormalization conditions that one imposes.

We stress that we shall integrate out only the KK continuum, first
of all because this is what we are interested in.  Furthermore, a
complete discussion of the effect of a very light mode would
require going to higher loops, and this is out of the scope of
this paper. According to the split of $\phisq$ made in Section
\ref{sec:bskk}, the contribution from the KK modes is given by
\beq%
\label{vkk}%
{\partial V^\kk \over \partial \varphi_c} \equiv 2 \lambda
\varphi_c\; {}^{(\kk)}\phisq|_0(\varphi_c) ~.
\eeq%

Because the 1-loop KK contribution depends only on $\varphi_c$,
the
total effective potential takes the form%
\beq\label{veff}%
V_\eff=V^{cl}\big(\varphi_c,\Psi_\bs\big)%
+V^\kk\big(\varphi_c\big)~,%
\eeq%
where $V^{cl}$ is given in  (\ref{vDR}). We shall impose the
following renormalization conditions
\begin{align}
\label{rencondkk} %
V^\kk|_{\varphi=0}&=0\cr
\partial_{\varphi_c}^2V^\kk|_{\varphi=0}&=0
\end{align}
The first condition ensures that in the $\varphi_c=0$ vacuum, the
cosmological constant is the same as in the background. The second
demands that $V_\eff''(\varphi)$  coincides with the mass at tree
level $m_\varphi^2$, at $\varphi_c=0$.

\begin{figure}[htb]
  \includegraphics[width=8cm]{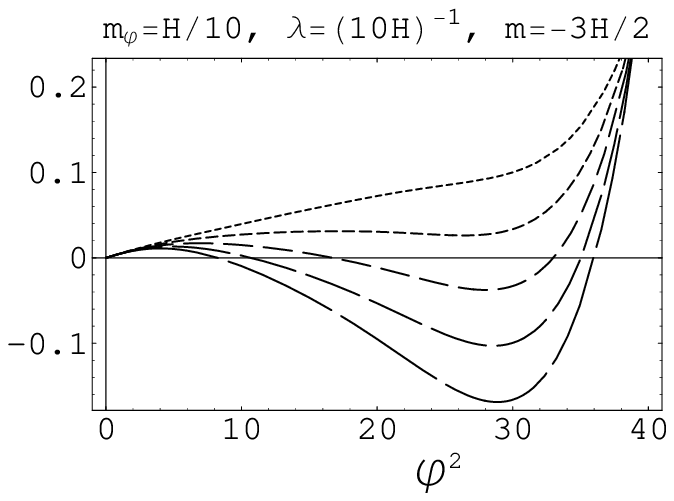}\\
  \caption{Form of the effective potential $V_\eff(\varphi)$
  for typical values of the parameters corresponding to the
  conformally coupled bulk field. We plot the difference with respect
  to the background vacuum energy density in units of
  $H^4$, and we represent it on the axis $\Psi_\bs=0$.
   The dotted line is with $\log(\mu/H)=-0.8$.
   For longer dashing, the logarithm is subsequently increased by
   $0.2$. We see how $V_\eff(\varphi)$ tends to
   acquire a true vacuum away from $\varphi=0$. Positive
   values of $\log(\mu/H)$ lead to more negative potential.
   }\label{fig:veffconf}
\end{figure}

We showed in Section \ref{sec:fluct} that $\phisq|_0$ is defined
up to the four constants $A$, $B$, $C$ and $\mu$ in
(\ref{result2}). The equations (\ref{rencondkk}) fixes one of
them. 
We will assume that the remaining renormalization constants are of
order one in the natural units of the problem. This leads to the
potential depicted in Fig. \ref{fig:veffconf} for natural choices
of the parameters. For $\lambda\varphi^2\ll H$, this potential can
be parametrized as
\beq\label{veffForm} %
V_\eff={1\over2}m_\varphi^2 \varphi^2 -{1\over4} a_4 \;H^2
(\lambda \varphi^2)^2 +
{1\over6} a_6 \;H (\lambda\varphi^2)^3 + \dots%
\eeq %
where $a_{4}$ and $a_{6}$ are numerical coefficients suppressed by
1-loop factors ({\em e.g.} $1/64\pi^2$, see Eq. (\ref{result2}))
and are typically of order $10^{-2}$--$10^{-3}$ . Physically, the
fluctuations ${}^{(\kk)}\phisq$ increase for $\varphi\to0$,
because in this limit there is a massless mode in the spectrum.
This implies a negative slope at $\varphi=0$ both in
${}^{(\kk)}\phisq$ and in $V^\kk$. That is why we take the
$\varphi^4$ term negative.
The extrema are located at%
\beq\label{extrema} %
\lambda\,\varphi^2={a_4 H\over2
a_6}\pm\sqrt{\lp({a_4H\over2a_6}\rp)^2- {m_\varphi^2\over \lambda
H\,a_6}},
\eeq %
the minus sign corresponding to the maximum, and the plus to the
new vacuum. This appears only for light enough $\varphi$ or
conversely when the interaction (\ref{int}) is strong enough,
\beq\label{condition} %
m_\varphi^2< \lambda H  {a_4^2\over 4a_6} H^2~.%
\eeq %

From (\ref{extrema}), we see that the location of the new vacuum
is such that
$$
\lambda\la\varphi\ra^2\gtrsim {a_4\over a_6} H~,
$$
and in the absence of fine tunings one expects this to exceed the
critical value $3H/2$. In this case, the parameter
$\nu=-\mbr-\lambda\varphi^2$ becomes negative, which renders the
bound state of $\Psi$ unnormalizable (see Eq. (\ref{bswf})). In
this situation, this mode becomes a quasi-normal mode with decay
width given by $\nu$, and it decays to the KK modes
\cite{rubakov}. Hence, the new minimum represents the true vacuum,
and the original one is at most meta-stable. From (\ref{veffForm})
that the value of the potential at the new minimum typically is
smaller than at the $\varphi=0$. At the new minimum, the quadratic
term in (\ref{veffForm}) can be neglected because it is only
comparable to the quartic term at the location of the maximum.
Hence, an order-of-magnitude estimate of the decrease in the
potential at the true vacuum is
$$
\delta V\sim -{1\over12}{a_4^3\over a_6^2}  H^4~,
$$
which is suppressed respect to $H^4$ by one 1-loop factor. This
gives a small correction to the background potential or
cosmological constant that is driving inflation, so a transition
to the true vacuum does not imply that inflation stops. It only
affects the spectrum of the bulk fields.

~\\

\section{Conclusions}
\label{sec:concl}

We have shown that the Kaluza-Klein excitations can considerably
modify the dynamics when interactions are included. We have
considered a bulk scalar field $\Psi$ coupled on the brane to a 4D
scalar field $\varphi$ with a bi-quadratic interaction on the
brane of the form (\ref{int}) taking as the background the RSII
space \cite{rsII} with an inflating brane. The bulk field $\Psi$
has an almost massless mode, the 'bound state'. We have computed
the effective potential $V_\eff(\varphi)$ induced by the KK modes.
The potential $V_\eff(\varphi)$ typically develops a minimum at a
value of $\varphi$ for which the bound state of $\Psi$ is no
longer normalizable. For natural choices of the renormalization
parameters, the potential in the new vacuum is smaller than in the
original one. This indicates that the vacuum $\varphi=0$ is
meta-stable.  In the true vacuum, the former bound state mode
becomes unstable-- it acquires a finite width and decays into bulk
modes \cite{rubakov}.

Intuitively, this happens because when the bound state of $\Psi$
is light, then the fluctuations $\phisq$ become large, as long as
the brane is inflating. Due to the interaction (\ref{int}),
$\phisq$ act as an effective mass term for $\varphi$. The
configuration with larger effective mass is disfavored, hence the
brane field $\varphi$ is driven away from $\varphi=0$ by the
quantum effects of the bulk field $\varphi$.

In this model, the KK modes seem to affect the dynamics
considerably. The reason is that in the one brane model with an
infinite extra dimension, the lightest KK mass is of order of the
Hubble constant, which is relatively light. Moreover, it is
natural that the fluctuations of the KK modes are sensitive to how
light is the lightest mode because, after all, they are part of
the same 5D field and the fluctuations have to increase in this
limit. Finally, the contribution to $\phisq$ from the bound state
is even larger than that of the KK modes (see Fig. \ref{fig:phi2},
and \cite{kks,shs}),
so if we include it then the instability is even stronger.\\

We also comment on a number of technical issues related to the
method used to obtain the 1-loop result for the quantum
fluctuations on the brane $\phisq|_0$ (\ref{result2}),  from which
we derive the effective potential $V_\eff(\varphi)$. One way to
compute $\phisq|_0$ is by taking into account the brane thickness,
where this feature is not expected to appear. In this article, we
have resorted to the thin wall approximation. A generic
consequence of this is that the vev of the field fluctuations
$\phisq$ blow up close to the brane. We showed that the field
fluctuations on the brane $\phisq|_0$ are well defined up to mass
and extrinsic curvature counter-terms.
Furthermore, we found that the renormalized values of fluctuations
on the brane $\phisq|_0$ and in the bulk $\phisq(z)$ (close enough
to the brane) coincide up to finite renormalization of mass and
extrinsic curvature counter-terms. This is what one would expect
to happen, and was previously noticed in \cite{ns} for vanishing
brane mass. The equivalence between $\phisq|_0$ and $\phisq(z)$ in
the less trivial case discussed here provides further evidence on
the validity of the thin wall approximation.\\

Our computation of $\phisq|_0$ shows that the KK continuum behaves
like a purely-decaying mode of a scalar field. For the range of
parameters that give rise to a normalizable bound state, the KK
mode contribution exactly cancels (up to finite renormalization of
mass terms) the decaying-mode contribution of the bound state. For
choices  of the parameters leading to no bound state, the KK
contribution behaves like the quasi-normal mode of the 5D field
$\Psi$, which is 'purely decaying' in the case of conformal
coupling. We leave for future investigation the analysis of bulk
scalar fields with generic mass and non-minimal coupling. Also,
our computation shows that the the bound state contribution
(\ref{bsalone}) to the field fluctuations on the brane $\phisq|_0$
dominate over the KK contribution (\ref{kk})) for $m_\bs\ll H$.
This agrees with the results of \cite{kks,hts}, where the
contribution to the power spectrum from the KK modes was found to
be negligible. One expects that this holds also for generic bulk
fields, though the analysis is slightly more technical and is left
for the future.

Our computation is relevant to the bulk inflaton model
\cite{hs,kks} where $\Psi$ plays the role of the a 5D field that
drives inflation. This field is assumed to have a light mode,
whose fluctuations seed the universe with the primordial
perturbations. Our result implies that the phase when $\Psi$ has a
light mode is limited by the instability of this vacuum. It seems
that the instability described here could places some constraints
on the model, which for instance depend on the mass of the brane
field $m_\varphi$ and the coupling constant $\lambda$. It is not
the purpose of this article to discuss the details of these
constraints, or the way how the decay of the false vacuum
proceeds, since they seem quite model-dependent. On the other
hand, it seems that this phenomenon can be extended to other brane
models. Whenever the brane is inflating, there is a bulk field
with a light mode and it is coupled to brane fields, the effective
potential should favor a vacuum where the fluctuations of the bulk
field are not so large, which precisely corresponds to not having
a very light mode.

\acknowledgements

We are grateful to Takahiro Tanaka, Jose Juan Blanco-Pillado,
Antonino Flachi, Jiro Soda, David Wands and Wade Naylor for useful
discussions. M.S. is supported in part by Monbukagakusho
Grant-in-Aid for Scientific Research (S) 14102004 and JSPS
Grants-in-Aid for Scientific Research (B) 17340075. O.P.
acknowledges support from JSPS Fellowship number P03193.

\appendix

\section{Conformal Transformations}
\label{app:conf}

The form of $\phisq$ for the flat ball (the interior of a
spherical cavity in flat space) was discussed in \cite{pt}. Using
the flat radial coordinate $\rho$, the metric for the flat ball is
\begin{equation}\label{flat}
ds_{\fl}^2=d\rho^2+\rho^2 \; dS_{(n+1)}^2,
\end{equation}
where $dS_{(n+1)}^2$ is the metric on a $n+1$ dimensional de
Sitter space. In terms of this coordinate, the renormalized Green
function for a conformally coupled field with no brane mass in the
de Sitter invariant vacuum is \cite{pt}
\begin{align}
\label{Gflat} %
G^{(1)}_\fl(x,x')&= {2\over n S_{(n+1)}}
\left({1\over \rho_0^2 + (\rho \rho'/\rho_0)^2 - 2 \rho \rho'
\cos\zeta} \right)^{{n \over 2}},
\end{align}
where $\rho_0$ is the location of the brane, $\zeta$ is the
geodesic distance in $dS$ space and $S_{(n+1)}=2
\pi^{n/2+1}/\Gamma(n/2+1)$ is the volume of a unit $n+1$
dimensional sphere. Equation (\ref{Gflat}) can be easily derived
by the method of images. The coincidence limit of this expression
is finite in the bulk, so we readily obtain
\begin{align}
\label{phi2flat} %
{}^{\fl}\phisq(x)= {1\over n S_{(n+1)} } \,%
\left({\rho_0\over \rho_0^2 - \rho^2 } \right)^{n}.
\end{align}

Now consider a bulk space of the form
\begin{equation}\label{ametric}
ds^2_{(a)}=dr^2+a^2(r)dS^2_{(n+1)}.
\end{equation}
Clearly, this is conformally related to flat space,
$$
ds^2_{(a)}=\Omega^2\,ds^2_{(0)},
$$
with $\Omega=a/\rho=\cosh^2\left(r/2\ell\right)$. Note that this
conformal factor is finite everywhere. The relationship between
the radial coordinate $r$ and the flat coordinate $\rho$ is
\begin{equation}
\label{rho} %
\ln\left(\rho_0/\rho\right)=\int_r^{r_0}{dr\over a(r)}\equiv z,
\end{equation}
and we have introduced the conformal coordinate $z$, in terms of
which the brane sits at $z=0$, and $r=\rho=0$ corresponds to
$z\to\infty$.

On the other hand, the Green function in the space (\ref{ametric})
for the conformally coupled scalar in the conformal vacuum is
\begin{equation}
  G^{(1)}_{(a)}(x,x')=  \left[\Omega(x)\Omega(x')\right]^{-n/2} \,G^{(1)}_{({\rm flat})}(x,x')
\end{equation}
Hence, in the space (\ref{ametric}) we obtain for this vacuum
\begin{align}
\label{phisq}
{}^{(a)}\phisq(x)= {a^{-n}\over n S_{(n+1)} } 
\left({\rho_0\,\rho\over \rho_0^2 - \rho^2 } \right)^{n}  ={\big(
2\, a\,\sinh{z}\big)^{-n}\over n S_{(n+1)} }~,
\end{align}
which agrees with \cite{ns}. This means that the procedure to
compute $\phisq$ used in \cite{ns}, based on a conformal
transformation to the cylinder adding a regulating brane and then
sending it to conformal infinity, works when computing local
quantities like $\phisq(z)$. Recently, it was shown in \cite{fkns}
that this procedure does not reproduce the correct results for
global quantities such as the effective action. The reason seems
to be that by introducing the second brane, one modifies the
topology (even in the limit when it is sent to infinity). However,
it seems reasonable that this procedure still works to compute
local quantities. Note that the method used in this Appendix also
makes use of a conformal transformation. However, it is perfectly
regular at all points, and the topology is not altered.
%

We can apply the same method to compute $G$ and $\phisq$ in the
case when we break conformal invariance in one point, on the
brane. The boundary condition in the original AdS space is
$$
[n^\mu\partial_\mu - 2\xi_c K - \mbr] \Psi|_0=0
$$
where $|_0$ denotes evaluation on the brane. 
Because of the mass term, this is not conformally invariant.
However, the conformal factor that links AdS with flat space is
constant on $r=$const. surfaces, so the boundary condition can be
written in the same form if we rescale the mass term as
$\mbr\to\mbr/\Omega_0$. Since the Hubble constant on the brane
$H=a_0^{-1}$ scales precisely in the same way, the parameter
$\nu=-\mbr/H$ (see Eq (\ref{nu})) does not scale. Hence, for
$M=0$, $\xi=3/16$ and $\mbr\neq0$, the form of $\phisq$ is the
same as (\ref{coinc}) for any space conformally related to the
flat ball.\\

Finally, we shall comment on one further check of (\ref{result2}).
In \cite{flatball}, the determinant of the Laplace operator for a
scalar field in the flat ball with Robin boundary conditions on
the boundary was computed (see \cite{kirsten} for the computation
in more general situations). This is equivalent to the effective
potential induced by a bulk field for any nonminimal coupling, as
a function of the boundary condition parameter. As discussed
above, this space is conformal to the AdS ball. The difference
between the effective potential of a given field in conformally
related spaces is known as the cocycle function, and can be
written in terms of the geometrical invariants of both spaces
\cite{gpt,gpt2,da}, through the Seeley-DeWitt coefficients. It
also depends on the field parameters (bulk and brane masses, etc).
It can be easily shown (see {\em e.g.} \cite{a5}) that the
dependence is polynomial in the boundary mass. Thus, the
dependence in $\mbr$ of the cocycle function connecting the flat
ball and the AdS ball reduces to pure counter-terms and can be
ignored. Indeed, it can easily be checked that the effective
potential found in \cite{flatball} satisfies (\ref{veff}) up to
polynomial terms in $\mbr$.

\end{document}